\documentclass[useAMS,usenatbib]{mn2e}
\usepackage{graphicx}

\def\f{\frac}

\def\BE{\begin{equation}}
\def\EE{\end{equation}}
\def\BEA{\begin{eqnarray}}
\def\EEA{\end{eqnarray}}

\def\d{{\rm d}}
\def\u{{\rm u}}

\def\max{{\rm max}}

\def\acc{{\rm acc}}

\def\MG{\mu{\rm G}}
\def\TeV{{\rm TeV}}

\def\epsilon{\varepsilon}

\def\gtrsim{\ga}
\def\lesssim{\la}
\def\apj{{ApJ}}
\def\aap{{A\&A}}
\def\pasj{{PASJ}}

\title[TeV gamma-rays from old SNRs]
{TeV Gamma-Rays from Old Supernova Remnants 
}
\author[R. Yamazaki et al.]
{Ryo Yamazaki,$^{1}$\thanks
{E-mail: ryo@theo.phys.sci.hiroshima-u.ac.jp (RY)}
Kazunori Kohri,$^{2}$
Aya Bamba,$^{3}$
Tatsuo Yoshida,$^{4}$
Toru Tsuribe$^{5}$
\newauthor
and
Fumio Takahara$^{5}$\\
$^{1}$Department of Physics, Hiroshima University, 
Higashi-Hiroshima, Hiroshima 739-8526,  Japan\\
$^{2}$Institute for Theory and Computation,
Harvard-Smithsonian Center for Astrophysics,
MS-51, 60 Garden Street, Cambridge, MA 02138, USA\\
$^{3}$RIKEN (The Institute of Physical and Chemical Research)
2-1, Hirosawa, Wako, Saitama 351-0198, Japan\\
$^{4}$Faculty of Science,
Ibaraki University, Mito 310-8512, Japan\\
$^{5}$Department of Earth and Space Science,
Osaka University, Toyonaka 560-0043, Japan
}
\begin{document}

\date{
Accepted 2006 July 18.  
Received 2006 July 12; 
in original form 2006 January 31 
}

\pagerange{\pageref{firstpage}--\pageref{lastpage}} \pubyear{2002}

\maketitle

\label{firstpage}

\begin{abstract}

We study
the emission from an old supernova remnant (SNR) with 
an  age of around $10^5$~yrs
and that from a giant molecular cloud (GMC)
encountered by the SNR.
When the SNR age is around $10^5$~yrs, 
proton acceleration is
efficient enough to emit TeV $\gamma$-rays 
both at  the shock of the SNR and that in the GMC.
The maximum energy of primarily accelerated electrons
is so small that TeV $\gamma$-rays and  X-rays
are dominated by hadronic processes,
 $\pi^0$-decay and synchrotron radiation from secondary
electrons, respectively.
However, if the SNR is older than several $10^5$~yrs,
there are few high-energy particles emitting TeV $\gamma$-rays
 because of the energy loss effect
and/or the wave damping effect occurring at low-velocity isothermal
shocks.
For old SNRs or SNR-GMC interacting systems capable of
generating TeV $\gamma$-ray emitting particles,
we calculated the
ratio of TeV $\gamma$-ray (1--10~TeV)  to X-ray (2--10~keV)
energy flux and found that it can be more than $\sim10^2$.
Such a source showing large flux ratio may be
a possible origin of recently discovered 
 unidentified TeV sources.
\end{abstract}

\begin{keywords}
acceleration of particles ---
gamma-rays: theory ---
ISM: clouds ---
shock waves --- 
supernova remnants ---  
X-rays: ISM
\end{keywords}

\section{Introduction}
\label{sec:intro}

The most probable cosmic-ray accelerator in our Galaxy is 
the young supernova remnant (SNR).
The detection of synchrotron X-rays from shells of young SNRs 
provides us the strong evidence for electron
acceleration up to more than $\sim10$~TeV
\citep[e.g.,][]{koyama1995,koyama1997,
bamba2003a,bamba2003b,bamba2005a,bamba2005b}.
So far, the evidence for hadron acceleration, however, has not
yet been obtained. High energy $\gamma$-ray observations may give 
us important information on the accelerated protons
\citep{naito1994,drury1994,aharonian1994,aharonian1996}. 
For example, TeV $\gamma$-rays are detected from the
SNRs RX~J1713.7$-$3946 \citep{enomoto2002,aharonian2004}
and RX~J0852.0$-$4622 \citep{katagiri2005,aharonian2005c},
which can be originated in either the
decay of neutral pion, arising from the collision of high energy
protons and interstellar matter, or CMB photons up-scattered
by accelerated electrons
\citep[e.g,][]{pannuti2003,lazendic2004,
ellison2001,bamba2005b,uchiyama2005}.
At present, the leptonic process is not yet ruled out.

Recently, a survey of the inner part of our Galaxy
has revealed several new TeV $\gamma$-ray sources 
\citep{aharonian2002,aharonian2005,aharonian2005b,aharonian2005d}. 
For some of them,
no counterpart has been found in any other wave lengths yet. 
Their spectra are rather hard, and
the flux above 1~TeV is around $1\times10^{-11}$~cm$^{-2}$s$^{-1}$.
They are extended with the angular size of around $0.1^\circ$.
They should be galactic origin because all are located along the
galactic plane.
In Table~\ref{table}, we list the properties of the
Galactic TeV $\gamma$-ray sources that are also observed in the X-ray bands
(together with a typical young SNR SN~1006).
One of the newly discovered sources, HESS~J1303$-$631 
\citep{aharonian2005b}, was
observed by {\it Chandra}, and no obvious counterpart was revealed
\citep{mukherjee2005}.
Then, the flux ratio, defined by
\begin{equation}
R_{\rm TeV/X}=\f{F_\gamma(1-10~{\rm TeV})}{F_X(2-10~{\rm keV})}~~,
\end{equation}
is more than $\sim2$.
On the other hand,  young SNRs, Cas~A, RX~J1713.7$-$3946 and RX~J0852.0$-$4622, 
have $R_{\rm TeV/X}$ less than $\sim2$.
Although at present the lower limit on $R_{\rm TeV/X}$ for 
HESS~J1303$-$631 is comparable to that of young SNRs, 
it may become much larger with
forthcoming deeper X-ray observations.
Furthermore, HESS~J1640$-$465 has an X-ray counterpart
that is identified as SNR \citep{aharonian2005d}.
This object has $R_{\rm TeV/X}\sim19$.
Unlike young SNRs, these TeV sources with large $R_{\rm TeV/X}$
 may show  evidence for
hadron acceleration because inverse-Compton scenario requires
unusually small magnetic field strength ($\ll1~\mu$G).
\begin{table*}
\centering
\begin{minipage}{140mm}
\caption{
Observed properties of Galactic TeV $\gamma$-ray sources and a 
typical young SNR, SN~1006.}
\label{table}
\begin{tabular}{cccccc}
\hline
Name & $F_\gamma$(1--10~TeV)%
\footnote{Derived for best fitted values (in units of $10^{-12}$erg~s$^{-1}$cm$^{-2}$).}
& $F_X$(2--10~keV)%
\footnote{Derived for best fitted values (in units of $10^{-12}$erg~s$^{-1}$cm$^{-2}$).}
& $R_{\rm TeV/X}$%
\footnote{$R_{\rm TeV/X}=F_\gamma$(1--10~TeV$)/F_X$(2--10~keV)~.}
& Classification%
\footnote{SNR: Supernova Remnant, PWN: Pulsar Wind Nebula,
unID: unidentified TeV $\gamma$-ray source.}
& References%
\footnote{
(1)~\citet{aharonian2004b};
(2)~\citet{willingale2001};
(3)~\citet{aharonian2005f};
(4)~\citet{delaney2006}
(5)~\citet{aharonian2001}; 
(6)~\citet{allen1997};
(7)~\citet{aharonian2005e}; 
(8)~\citet{ozaki1998};
(9)~\citet{aharonian2004}; 
(10)~\citet{pannuti2003};
(11)~\citet{aharonian2005c}; 
(12)~\citet{slane2001};
(13)~\citet{aharonian2005d}; 
(14)~\citet{brogan2005};
(15)~\citet{sugizaki2001};
(16)~\citet{aharonian2002}; 
(17)~\citet{aharonian2005b};
(18)~\citet{mukherjee2005}.
}
\\
\hline
Crab              & $56$   & $2.1\times10^{4}$  & $2.6\times10^{-3}$ & PWN  & (1), (2)  \\
RCW~89            & $16$   & $38$             & $0.41$  & PWN  & (3), (4)  \\
Cas~A             & $1.9$  & $1.3\times10^2$  & $0.014$ & SNR  & (5), (6)  \\
SN~1006 (NE)      & $<2.1$ & $19$             & $<0.11$ & SNR  & (7), (8)  \\
RX~J1713.7$-$3946 & $35$   & $5\times10^2$    & $0.07$  & SNR  & (9), (10)  \\
RX~J0852.0$-$4622 & $69$   & $>32$            & $<2.1$  & SNR  & (11), (12)  \\
HESS~J1813$-$178  & $8.9$  & $7.0$            & $1.3$   & SNR  & (13), (14) \\
HESS~J1640$-$465  & $7.1$  & $0.37$           & $19$    & SNR  & (13), (15) \\
TeV~J2032$+$4130  & $1.9$  & $<2$             & $>1$    & unID & (16)       \\
HESS~J1303$-$631  & $10$   & $<6$             & $>2$    & unID & (17), (18) \\
\hline
\end{tabular}
\end{minipage}
\end{table*}

A large value of $R_{\rm TeV/X}$ is expected
for old SNRs.
As the SNR ages,  the shock velocity decreases. 
In general, primary electron acceleration is limited by synchrotron cooling. 
Then, the roll-off energy of electron synchrotron radiation
is much smaller than that of typical young SNRs,
so that  small synchrotron X-ray flux is expected 
\citep[e.g.,][]{sturner1997}.
It is also important to consider the association with a
giant molecular cloud (GMC).
Because of large volume, old SNRs may encounter the GMC.
In this paper, we study
how large $R_{\rm TeV/X}$ can become for
 single SNRs and the SNR-GMC interacting systems.

\section{Evolution of SNR}
\label{sec:evolution}

We consider a simple analytical model of
 the shock dynamics of SNRs expanding into  the uniform
ambient medium with the density $n_0$.
The SNRs evolve through three phases:
the free expansion phase, the Sedov-Taylor phase,
and the radiative phase.
We assume the shock velocity $v_s$ as a function of the SNR age
$t_{\rm age}$ as
\begin{equation}
v_s(t_{\rm age})=
\left\{
\begin{array}
{l@{} l@{}}
v_i & (0<t_{\rm age}<t_1) \\
v_i (t_{\rm age}/t_1)^{-3/5} & (t_1<t_{\rm age}<t_2) \\
v_i (t_2/t_1)^{-3/5}(t_{\rm age}/t_2)^{-2/3}~ & 
(t_2<t_{\rm age})
\end{array} \right.~~ ,
\label{eq:Vs}
\end{equation}
where 
$t_1=(3E/2\pi m_Hn_0v_i^5)^{1/3}
=2.1\times10^2(E_{51}/n_0)^{1/3}v_{i,9}^{-5/3}$~yrs
and 
$t_2=4\times10^4E_{51}^{4/17}n_0^{-9/17}$~yrs \citep{blondin1998},
and $v_i=v_{i,9}10^9$~cm~s$^{-1}$ and $E=E_{51}10^{51}$~ergs are 
the initial velocity,
and the initial energy of the ejecta, respectively.
Here we adopt the expansion law of the radiative phase,
$v_s\propto t_{\rm age}^{-2/3}$, instead of 
$v_s\propto t_{\rm age}^{-5/7}$, because 
the former gives a better approximation
in the epoch around $t_{\rm age}\sim10^5$~yrs
\citep{blondin1998,bandiera2004}.
The evolution of the shock radius $R_s=\int v_s dt$ 
is also calculated.
In this paper,  we adopt $E_{51}=v_{i,9}=n_0=1$
as a fiducial case. Then,
we find $R_s=91$~pc and $v_s=1.1\times10^7$~cm~s$^{-1}$
at $t_{\rm age}=3\times10^5$~yrs. 
Note that the effect of  the magnetic field on SNR dynamics
can be neglected until $t_{\rm age}$ of
several $10^5$~yrs \citep[e.g.,][]{hanayama2006}.

\section{Emission from an SNR}
\label{sec:emissionSNR}

At first, we consider the spectrum of emitting high-energy particles
that get their energy via diffusive shock acceleration
\citep{drury1983,blandford1987}.
Generally speaking, in order to  obtain the energy spectrum
of accelerated particles,  time-dependent kinetic equation 
should be solved
as investigated by many authors \citep[e.g.,][]{berezhko1996},
even including the nonlinear effects via accelerated particles
\citep{malkov2001}.
Such  calculations  well match  the observational facts
when free parameters such as 
the injection rate and  the magnetic field
are chosen appropriately.
Instead, we assume, in this paper, that at an arbitrary epoch, the
spectral form is a  power-law with the index $p$
and the exponential cut-off at $E_{\rm max}$, i.e.,
$\propto E^{-p}e^{-E/E_{\rm max}}$, 
where $E_{\rm max}$ evolving with time.
This approximation may be valid and be useful to extract
the basic properties of high-energy emission from a SNR because of 
the following reasons.
High-energy particles are produced by the diffusive shock acceleration
and suffer adiabatic expansion after they are transported
downstream of the shock and lose their energy. 
Hence, at the given epoch, the spectrum for high-energy particles is
dominated by those which are being accelerated at that time, in other
words, the energy spectrum of particles does not so much depend
on the past acceleration history.
Especially, we are now interested in the energy region near the upper end
of the spectrum because as seen in the following,
both X-rays and TeV $\gamma$-rays are produced
by particles with energy near $E_{\rm max}$.
In  this energy regime, our assumed form of the spectrum
may be a good approximation.

Once the  SNR dynamics is given,
the maximum energy of accelerated particles is calculated.
The maximum energy of accelerated protons, $E_{\rm max,p}$, is determined by
\begin{equation}
t_{\rm acc}={\rm min}\{t_{\rm age},t_{\rm pp}\}~~,
\label{eq:Tacc_p}
\end{equation}
while that of accelerated electrons, $E_{\rm max,e}$, is determined by
\begin{equation}
t_{\rm acc}={\rm min}\{t_{\rm age},t_{\rm synch}\}~~,
\label{eq:Tacc_e}
\end{equation}
where $t_{\rm age}$, $t_{\rm acc}$,  $t_{\rm synch}$, and $t_{\rm pp}$
are the age of the SNR, the acceleration time scale, 
the synchrotron loss time scale, and the pion-production loss time scale,
respectively.
Since we assume that the  energy spectrum does not depend on
the past history,
 it is also a good approximation to determine $E_{\rm max}$
using Eqs.~(\ref{eq:Tacc_p}) or (\ref{eq:Tacc_e}).
Indeed, at least for young SNRs, one can confirm 
 our estimation of $E_{\rm max}$ coincides with
 simulation results \citep{berezhko2002}.
For the diffusive shock acceleration,
the acceleration time is given as
\begin{equation}
t_\acc=\f{20hcE_\max}{eB_\d v_s^2}~~,
\end{equation}
where 
\begin{equation}
h=\f{0.05r(f+rg)}{r-1}~~, 
\end{equation}
and $r$ is the compression ratio, and
 $f$ and $g$  are functions of the
shock angle $\theta$ and gyro-factors $\eta_\u$ and $\eta_\d$
that are given as 
\begin{equation}
f(\eta_\u,\theta)=
\eta_\u (\cos^2\theta+r^2\sin^2\theta)^{1/2}
\left(\cos^2\theta+\f{\sin^2\theta}
{1+\eta_\u^2}\right)\  ,
\label{DefF}
\end{equation}
\begin{equation}
g(\eta_\d,\theta)=
\eta_d (\cos^2\theta+r^2\sin^2\theta)^{-1}
\left(\cos^2\theta+\f{r^2\sin^2\theta}
{1+\eta_\d^2}\right) \ ,
\label{DefG}
\end{equation}
respectively \citep{yamazaki2004,jokipii1987}.
The downstream magnetic field is given by $B_\d=rB_{\rm ISM}$,
where $B_{\rm ISM}=10B_{\rm ISM,-5}~\MG$ is the ISM magnetic field
and we adopt $B_{\rm ISM,-5}=1$ as a typical value.
For high-energy protons, the energy loss time scale through
pion-production is given by
$t_{\rm pp}= 5.3\times10^7n^{-1}$~yrs, where $n$ is the number density
of the acceleration site.
For electrons, the synchrotron loss time scale is given by
$t_{\rm synch}=1.25\times10^4(E_{\rm max,e}/10{\rm TeV})^{-1}(B_\d/10\MG)^{-2}$~yrs.
As long as  $t_{\rm age}<t_{\rm pp}$,
 the proton acceleration is age-limited, and we obtain
\begin{equation}
E_{\rm max,p}=1.6\times10^2~{\rm TeV}~h^{-1}v_{s,8}^2
\left(\frac{B_{\rm d}}{10~\MG}\right)
\left(\frac{t_{\rm age}}{10^5{\rm yrs}}\right)~~,
\label{eq:Emax_p}
\end{equation}
where $v_{s,8}=v_s/10^8$cm~s$^{-1}$.
The electron acceleration is loss-limited
when $t_{\rm age}\gtrsim10^3$~yrs, and then we derive
\begin{equation}
E_{\rm max,e}=14~{\rm TeV}~h^{-1/2}v_{s,8}
\left(\frac{B_{\rm d}}{10\MG}\right)^{-1/2}~~.
\label{eq:Emax_e}
\end{equation}
The minimum energy, $E_{\rm min,e}$ and $E_{\rm min,p}$ are simply 
taken as the rest-mass energy of electrons and protons, respectively.

The wide-band radiation spectrum is calculated
for  given $E_{\rm max,p}$ and $E_{\rm max,e}$ at a certain
$t_{\rm age}$.
We consider  radiation processes from primary electrons:
the synchrotron, inverse-Compton (IC), and bremsstrahlung
emissions, and  from primary protons:
$\pi^0$ decay $\gamma$-rays, and synchrotron and bremsstrahlung
emissions from
secondary electrons arising from the decay of charged pions.
We assume the electron-proton ratio,
that is the ratio of  the electron distribution function to that
of protons for a fixed energy in relativistic regimes,
of $K_{\rm ep}=1\times10^{-3}$.
For average cosmic-rays in our Galaxy, $K_{\rm ep}$ is about $10^{-2}$,
which is an order of magnitude larger than our adopted value.
 However, in general the average value  may be
different from that at the acceleration site because of the propagation
effect. Indeed, as seen in the following,
if we adopt $K_{\rm ep}=1\times10^{-3}$, theoretically calculated
 flux ratio $R_{\rm TeV/X}$ for a young SNR is similar to the observed one.
For old SNRs, which we are most interested in,
the result for $R_{\rm TeV/X}$ is unchanged even if  $K_{\rm ep}$ varies
more than one order of magnitude because accelerated electrons do not contribute
to TeV  and X-ray emission.

Primarily accelerated electrons produce synchrotron emission with 
the roll-off frequency given by
$\nu_{\rm roll}\sim1.6\times10^{16}
(B_\d/10\MG)(E_{\rm max,e}/10\TeV)^2$~Hz
\citep{reynolds1999}\footnote{
See http://heasarc.gsfc.nasa.gov/docs/xanadu/xspec \\
for the erratum of coefficients.}.
Using Eq.~(\ref{eq:Emax_e}),
$\nu_{\rm roll}$ can be rewritten as
\begin{equation}
\nu_{\rm roll}\sim3.2\times10^{16}h^{-1}v_{s,8}^2~{\rm Hz}~~,
\label{eq:rolloff}
\end{equation}
in the loss-limited case \citep[see also][]{aharonian1999}.
Note that $\nu_{\rm roll}$ does not depend on $B_{\rm ISM}$.

\begin{figure}
\begin{center}
\includegraphics[width=80mm]{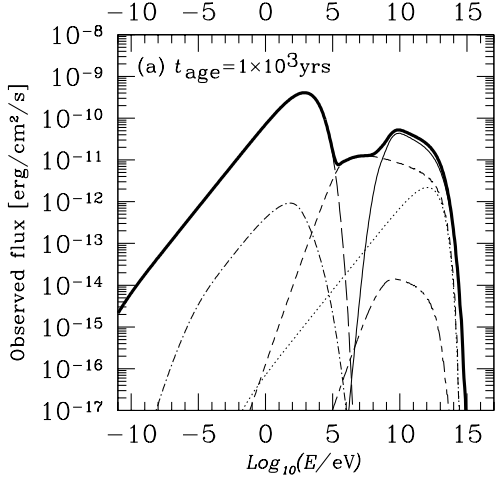}
\end{center}
\caption{
$\nu F_\nu$  spectrum of  a single SNR with an age
$1\times10^3$~yrs that stores energy,
$10^{50}$~ergs, of high-energy protons.
The thick-solid line shows the total non-thermal flux.
Hadronic emissions are 
$\pi^0$-decay $\gamma$-rays (thin-solid),
synchrotron (dot-dashed) and
bremsstrahlung emission (short-and-long dashed)
from secondary electrons produced by charged pion.
Leptonic emissions are
synchrotron (long-dashed),
inverse-Compton (dotted)
and bremsstrahlung (short-dashed) emission by
primary electrons.
The source is located at 1~kpc.
We adopt $E_{51}=v_{i,9}=n_0=h=B_{\rm ISM,-5}=1$, $p=2.2$, $r=4$ and
$K_{\rm ep}=1\times10^{-3}$  
(see text for details).
}
\label{fig:spec_snr_young1}
\end{figure}

\begin{figure}
\begin{center}
\includegraphics[width=80mm]{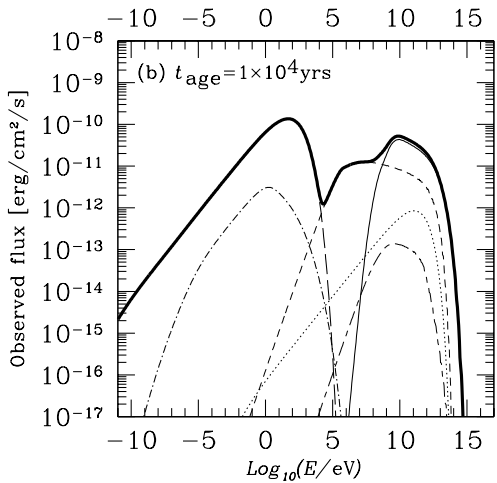}
\end{center}
\caption{
The same as Fig.~\ref{fig:spec_snr_young1} but for
an SNR age of $1\times10^4$~yrs.
}
\label{fig:spec_snr_young2}
\end{figure}

\subsection{Emission from a young SNR in the Sedov phase}
\label{subsec:emissionSNRyoung}

As can be seen from Eqs.~(\ref{eq:Emax_p}) and (\ref{eq:Emax_e}),
once the SNR dynamics and $B_{\rm ISM}(=B_\d/r)$ are fixed,
only necessary are the values of $r$ and $h$ in order to
calculate $E_{\rm max,p}$ and $E_{\rm max,e}$ for a given time.
In the Sedov phase
($t_1=2\times10^2~{\rm yrs}<t_{\rm age}<t_2=4\times10^4~{\rm yrs}$
for $E_{51}=n_0=v_{i,9}=1$),
the shock of an SNR is strong and adiabatic, 
so that the density compression ratio, $r$, is near 4.
Then in the Bohm limit case, $\eta_\u\sim\eta_\d\sim1$,
we find $h\sim1$ for arbitrary $\theta$.
When the nonlinear effects are considered, $r$ becomes as large as
 $\sim7$ \citep{berezhko2002}, 
however, in this case, one can find that
$h$ weakly depends on  $r$. 
Hence in this paper,
we neglect the nonlinear effects for simplicity.
When  $h$ is around unity,
we can reproduce the observed value of $\nu_{\rm roll}$  for young
SNRs \citep[e.g.,][]{bamba2003b,bamba2005a,bamba2005b}.
Hence,  for $t_{\rm age}\lesssim10^4$~yrs,
we adopt $r=4$ and $h=1$ as a fiducial value.
The power-law index of accelerated particles is fixed as $p=2.2$,
which is typical for young SNRs.

Here we assume that an
SNR stores energy, $10^{50}$~ergs, of high-energy protons.
As long as accelerated particles maintains upstream turbulence
via streaming instability, the self-confinement of accelerated 
particles is efficient, so that the escape of them upstream
may be neglected. Furthermore, radiative efficiency is also
small. As discussed previously, the energy of accelerated particles
is reconverted to the expansion energy through the adiabatic expansion,
which implies the expansion energy remains almost unchanged.
 Therefore, if the injection rate is time-independent,
the energy of high-energy protons is also time-independent.

When $t_{\rm age}=1\times10^3$~yrs
(then, $t_1<t_{\rm age}<t_2$),
we find $E_{\rm max,p}=96$~TeV, $E_{\rm max,e}=27$~TeV and
$\nu_{\rm roll}=4.8\times10^{17}$~Hz for the fiducial parameters, 
and find that nonthermal X-rays are dominated by
the synchrotron emission from primary electrons
(see Fig.~\ref{fig:spec_snr_young1}).
TeV $\gamma$-rays are dominated by $\pi^0$-decay process.
The flux ratio is $R_{\rm TeV/X}=7.6\times10^{-2}$,
 which is consistent with observations for young SNRs.

When $t_{\rm age}=1\times10^4$~yrs (Fig.~\ref{fig:spec_snr_young2}),
an SNR is still in the Sedov phase ($t_1<t_{\rm age}<t_2$).
For the fiducial parameters,
$E_{\rm max,p}=61$~TeV  and $E_{\rm max,e}=6.9$~TeV are derived.
Also in this case, the X-ray band is dominated by 
the primary synchrotron radiation.
As the SNR ages, $\nu_{\rm roll}$ becomes small, so that 
synchrotron radiation flux in the X-ray band becomes small. 
Hence the flux ratio becomes large, $R_{\rm TeV/X}=1.6$.

\begin{figure}
\begin{center}
\includegraphics[width=80mm]{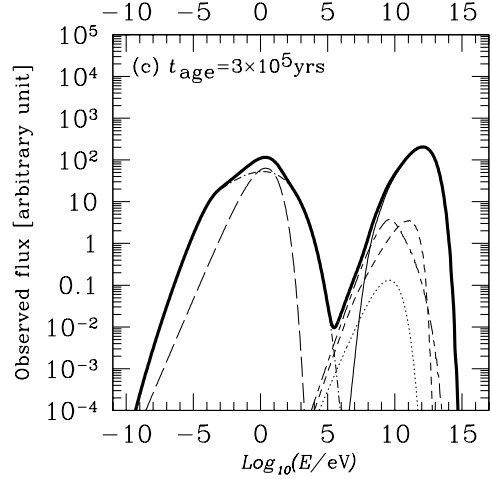}
\end{center}
\caption{
$\nu F_\nu$  spectrum of  a single SNR with an age
$3\times10^5$~yrs.
Meanings of each lines are the same as those in
Fig.~\ref{fig:spec_snr_young1}.
The absolute value of the observed flux is uncertain,
hence the flux is arbitrary scaled.
We adopt $E_{51}=v_{i,9}=n_0=h=B_{\rm ISM,-5}=1$, $p=1.5$, and
$K_{\rm ep}=1\times10^{-3}$  
(see text for details).
}
\label{fig:spec_snr_old1}
\end{figure}

\subsection{Emission from an old SNR in the radiative phase}
\label{subsec:emissionSNRold}

When an SNR enters into the radiative phase ($t_{\rm age}>t_2$),
the cooling effect becomes important, and 
the accumulated gas forms a dense, cool shell with
the number density $n_{\rm sh}=rn_0$.
Rewriting Eq.~(\ref{eq:Vs}) as
\begin{equation}
v_s=2.3\times10^7E_{51}^{11/51}n_0^{-4/17}
\left(\frac{t_{\rm age}}{10^5{\rm yrs}}\right)^{-2/3}
~{\rm cm}~{\rm s}^{-1}~~,
\label{eq:Vs_radiative}
\end{equation}
we find $v_s\sim10^7$~cm~s$^{-1}$ when $t_{\rm age}\sim10^5$~yrs.
Note that $v_s$ is independent of $v_i$.

In the radiative phase, the shock is isothermal.
Hence there is a concern that since the downstream temperature is low,
 the neutral component appears, and Alfv\'{e}n wave turbulence that scatters
nonthermal particles is significantly suppressed via the neutral-ion
friction, making the acceleration process inefficient
\citep{drury1996,bykov2000}. 
On the other hand, when the gas remains fully ionized around the shock front,
diffusive shock acceleration works well in the radiative phase.
\citet{shull1979} found that as long as 
$v_{s,7}\gtrsim1.1$, where $v_{s,7}=v_s/10^7$cm~s$^{-1}$, 
UV flux radiated in the
downstream region is sufficient enough to keep upstream state
fully ionized.
For $n_0=E_{51}=1$, we find $v_{s,7}=1.1$ at
$t_{\rm age}=3\times10^5$~yrs.
Therefore, when $t_{\rm age}\lesssim3\times10^5$~yrs, 
diffusive shock acceleration works well, and we can
use Eqs.~(\ref{eq:Emax_p}) and (\ref{eq:Emax_e}) in order to
estimate the maximum energy of accelerated particles.

The estimation of downstream magnetic field, $B_\d$, which appears
in Eqs.~(\ref{eq:Emax_p}) and (\ref{eq:Emax_e}), is different
from that for the young SNRs.
For $n_0=B_{\rm ISM,-5}=1$, the magnetic pressure
is larger than the gas pressure in the upstream region.
Then, solving the shock jump conditions 
for the quasi-perpendicular ($\theta\sim90^\circ$) isothermal shock,
the compression ratio is  given by 
\begin{equation}
r\sim\sqrt{2}\f{v_s}{v_{A1}}=6.5~B_{\rm ISM,-5}{}^{-1}
n_ 0^{1/2}v_{s,7}~~,
\end{equation}
where $v_{A1}=B_{\rm ISM}(4\pi m_Hn_0)^{-1/2}$
is the upstream Alfv\'{e}n velocity \citep{spitzer1978}. 
Therefore, the downstream magnetic field 
$B_\d=rB_{\rm ISM}=(8\pi m_Hn_0)^{1/2}v_s$ is
calculated as
\begin{equation}
B_\d=65~\MG~n_0^{1/2}v_{s,7}~~.
\label{eq:Bd_SNR} 
\end{equation}
Recently, using two-dimensional MHD simulation, 
\citet{hanayama2006} have shown
that in the case of $n_0=0.2$ (or 1) and 
$E_{51}=B_{\rm ISM}=0.5$,
the compression ratio at the quasi-perpendicular shock region
is about 7 at $t_{\rm age}$ of several $10^5$~yrs, 
which is roughly consistent with our estimation.
Using Eqs.~(\ref{eq:Emax_p}), (\ref{eq:Emax_e}), 
(\ref{eq:Vs_radiative}) and  (\ref{eq:Bd_SNR}),
we obtain
\begin{equation}
E_{\rm max,p}=34~{\rm TeV}~h^{-1}E_{51}^{11/34}n_0^{-6/17}
v_{s,7}^{3/2}~~,
\label{eq:Emax_p_old}
\end{equation}
\begin{equation}
E_{\rm max,e}=0.55~{\rm TeV}~h^{-1/2}n_0^{-1/4}
v_{s,7}^{1/2}~~.
\label{eq:Emax_e_old}
\end{equation}
Note that Eq.~(\ref{eq:Bd_SNR}) is derived for the limiting case
of a small upstream gas pressure compared to the magnetic pressure there.
If the explosion occurs where the gas pressure is comparable to
or larger than the magnetic pressure,  
 the effect of the gas pressure should be taken into account.
Then, the compression ratio becomes large \citep{spitzer1978}. 
Furthermore, when the accelerated protons are stored
around the shock front, a part of their energy goes into
the magnetic field energy, causing the magnetic field amplification
\citep{lucek2000}.
Therefore,  Eq.~(\ref{eq:Bd_SNR}) gives, in fact,
the lower bound of $B_\d$, so that,
Eqs.~(\ref{eq:Emax_p_old}) and (\ref{eq:Emax_e_old}) give 
lower and upper bounds of $E_{\rm max,p}(\propto B_\d)$ and 
$E_{\rm max,e}(\propto B_\d{}^{-1/2})$,
respectively.
In the following, however, 
neglecting such effects for simplicity, we use
Eqs.~(\ref{eq:Bd_SNR}), (\ref{eq:Emax_p_old}) and (\ref{eq:Emax_e_old})
to determine $B_\d$, $E_{\rm max,p}$ and $E_{\rm max,e}$.
One should also remark
 that Eq.~(\ref{eq:Bd_SNR}) is derived for
the quasi-perpendicular shock  ($\theta\sim90^\circ$).
Then, one can consider the acceleration at
the quasi-perpendicular shock as discussed in \citet{jokipii1987}.
However, even if $\theta\sim0$ upstream, then
 the accelerated particles penetrate into upstream region
and generate waves via streaming instability, resulting
in the generation of  oblique or quasi-perpendicular magnetic 
field upstream,
because the magnetic field component of generated waves is
generally  perpendicular to the shock normal and
can be as large as the original upstream magnetic field.

When $t_{\rm age}=3\times10^5$~yrs, we find
$r=B_\d/B_{\rm ISM}\sim7$.
Hence, in the case of Bohm limit $\eta_\u\sim\eta_\d\sim1$,
we find $h\sim1$ as well as for the young SNRs.
Until this epoch, the particle acceleration proceeds, so that
the upstream turbulence is maintained and the Bohm limit 
diffusion can be expected.
Therefore, we again take $h=1$ as a fiducial value.
In the test particle approximation, 
the power-law index of accelerated particles
$p$ and the compression ratio $r$ are
 related as $p=(r+2)/(r-1)$ \citep{blandford1987}.
If we take $r\sim7$ as a typical value, $p$ becomes about 1.5.
Hence, we adopt $p=1.5$ as a fiducial value.
However, as can be seen in the following,
our conclusion does not depend on $p$.

The total energy of accelerated protons stored in the SNR is
somewhat uncertain. 
If the confinement of accelerated particles works well
until $t_{\rm age}\sim10^5$~yrs,
the energy of accelerated protons may not be so small
compared with the young SNR case.
Otherwise, it is much smaller than $10^{50}$~ergs.
In this paper, since we are mainly 
interested in the value of $R_{\rm TeV/X}$,
 the total energy of accelerated particles, which
only determines the normalization of the radiated spectra,
is not essential for our arguments.
Further discussion on determining the
amount of accelerated particles will be seen in
\S~\ref{sec:dis}.

Figure~\ref{fig:spec_snr_old1} shows the spectrum 
for $t_{\rm age}=3\times10^5$~yrs
with fiducial parameters (i.e., $B_\d=72~\MG$,
$E_{\rm max,p}=42$~TeV and $E_{\rm max,e}=0.58$~TeV).
Then the flux ratio is $R_{\rm TeV/X}=82$.
Since $E_{\rm max,e}$ is  small,
TeV $\gamma$-rays via IC and bremsstrahlung are suppressed, and
therefore, the TeV $\gamma$-rays come from the $\pi^0$-decay.
The roll-off frequency of synchrotron radiation from
primary electrons is so small, $\nu_{\rm roll}\sim4.0\times10^{14}$~Hz,
that the secondary synchrotron radiation dominates the  X-ray band.
Also in this subsection, we have assumed the electron-to-proton
ratio as $K_{\rm ep}=1\times10^{-3}$.
However, even if we adopt much larger $K_{\rm ep}$,
$R_{\rm TeV/X}$ remains unchanged because as can be seen
in Fig.~\ref{fig:spec_snr_old1}, both X-ray and TeV $\gamma$-ray
bands lie above the cutoff of emission components
originating in primary electrons.

\begin{figure}
\begin{center}
   \includegraphics[width=80mm]{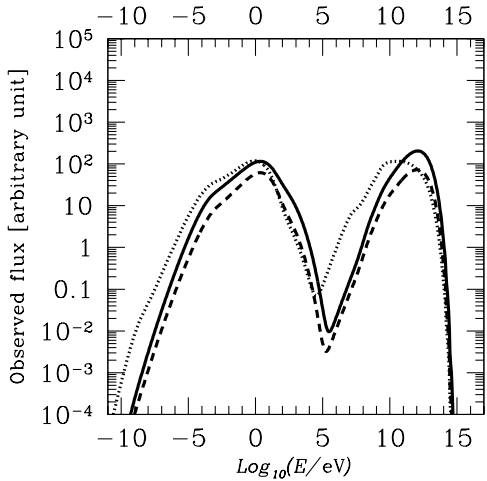}
\end{center}
\caption{
$\nu F_\nu$  spectrum of  a single SNR with an age
$3\times10^5$~yrs for various parameter sets.
Solid line is the total flux for the fiducial set
of parameters
($n_0=1$, $p=1.5$, 
$B_\d=72$~$\MG$, $E_{\rm max,p}=42$~TeV and $E_{\rm max,e}=0.58$~TeV), 
and is the same as the thick solid line
in Fig.~\ref{fig:spec_snr_old1}.
Dashed and dotted lines are for $n_0=0.5$ 
($B_\d=51$~$\MG$, $E_{\rm max,p}=39$~TeV and 
$E_{\rm max,e}=0.70$~TeV) and $p=2$ with the other parameters
 fiducial, respectively. 
}
\label{fig:spec_old2}
\end{figure}

We discuss how  $R_{\rm TeV/X}$ varies
for different values of parameters.
At first, changes of $B_{\rm ISM}$ and $v_{i}$ do not affect 
the conclusion because Eqs.~(\ref{eq:Bd_SNR}),
(\ref{eq:Emax_p_old}) and (\ref{eq:Emax_e_old}) tells us that
$B_\d$, $E_{\rm max,p}$ and $E_{\rm max,e}$, which determine
the overall shape of the emission spectrum, do not depend on
them.
As stated above, uncertainty of $K_{\rm ep}$ does not affect
the value of $R_{\rm TeV/X}$.
We also find  that $R_{\rm TeV/X}$ only weakly depends on
$h$ and $E_{51}$ for reasonable parameter ranges.
Hence we discuss the dependence of $n_0$ and $p$ in the following.
Let us first discuss the case of $n_0=0.5$ with other parameters 
unchanged.
We consider the epoch in which $v_{s,7}\sim1.1$
($t_{\rm age}\sim3.8\times10^5$~yrs for our parameters)
 when
$R_{\rm TeV/X}$ becomes maximum since it increases with time.
Then, we find $B_\d=51$~$\MG$, $E_{\rm max,p}=39$~TeV and 
$E_{\rm max,e}=0.70$~TeV, respectively.
Again both the X-rays and TeV $\gamma$-rays are dominated 
by the hadronic emission.
Compared with the fiducial case, the secondary synchrotron 
X-ray emission is dim because $E_{\rm max,p}$ and $B_\d$ are small
(see the dashed line in Fig.~\ref{fig:spec_old2}), 
so that the flux ratio becomes large
$R_{\rm TeV/X}=1.3\times10^2$.
Next,  we consider the case $p=2$ with other parameters
being fiducial (see the dotted line in Fig.~\ref{fig:spec_old2}).
TeV $\gamma$-rays  are produced by protons with energies of
$\sim10$~TeV, while X-rays come from secondary electrons
with energies of $\sim30$~TeV originated in $\sim100$~TeV protons.
Then, for a large $p$,  there are less protons with $\sim100$~TeV 
and the dimmer X-rays. Hence we find  the flux ratio becomes as large
as $R_{\rm TeV/X}=1.6\times10^2$.
So one can see that when parameters are changed within a 
reasonable range, the value of $R_{\rm TeV/X}$ remains unchanged
within a factor of two or three.

If $t_{\rm age}\gtrsim3\times10^5$~yrs, 
i.e., $v_{s,7}\lesssim1.1$, then, 
the ionization around the shock front is incomplete \citep{shull1979} 
and upstream ion-neutral Alfv\'{e}n wave damping 
places significant restriction on shock acceleration
\citep{drury1996,bykov2000}.  
Once the shock acceleration becomes inefficient, there are
few high-energy protons emitting TeV $\gamma$-rays
around the SNR shell, because
they escape the SNR shell due to the diffusion;
assuming the Bohm diffusion, the escape time for
a particle with an energy $E_{\rm cr}=10E_{\rm cr, 10TeV}$~TeV 
is estimated as
$t_{\rm esc}\sim1\times10^5\eta^{-1}(B_\d/70\MG)
E_{\rm cr, 10TeV}^{-1}\Delta_{\rm 3pc}{}^2$~yrs,
where $\eta$ and $\Delta=3\Delta_{\rm 3pc}$~pc are
the gyro factor and the thickness of the shell,
respectively.
Therefore, when $t_{\rm age}\gtrsim3\times10^5$yrs, 
TeV $\gamma$-rays are significantly suppressed.

\section{Emission from an old SNR interacting with a GMC}
\label{sec:emissionGMC}

\begin{figure}
\begin{center}
\includegraphics[width=80mm]{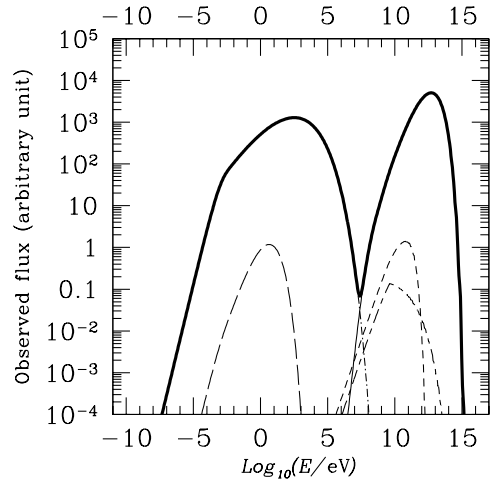}
\end{center}
\caption{
$\nu F_\nu$  spectrum for emission from a shocked GMC
that is collided by a SNR with an age of $4.6\times10^4$~yrs.
The GMC has mass of $10^5M_{\sun}$, and
the meanings of each line are the same as those in 
Fig.~\ref{fig:spec_snr_young1}. 
The flux is arbitrarily scaled.
We adopt $E_{51}=v_{i,9}=n_0=h=B_{\rm ISM,-5}=M_5=1$,
$K_{\rm ep}=1\times10^{-3}$ and $p=1$
(see text for details).
}
\label{fig:spec_gmc}
\end{figure}

\begin{figure}
\begin{center}
\includegraphics[width=80mm]{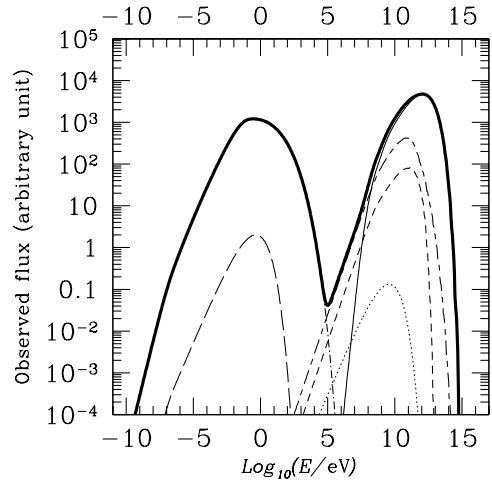}
\end{center}
\caption{
$\nu F_\nu$ emission spectrum from a GMC illuminated by 
high-energy particles
arising from the SNR with an age of  $3\times10^5$~yrs.
The GMC has mass of $10^5M_{\sun}$, and
the meanings of each line are the same as those in 
Fig.~\ref{fig:spec_snr_young1}. 
The flux is arbitrarily scaled.
We adopt $E_{51}=v_{i,9}=n_0=h=B_{\rm ISM,-5}=M_5=1$,
$K_{\rm ep}=1\times10^{-3}$ and $p=1.5$
(see text for details).
}
\label{fig:spec_gmc2}
\end{figure}

One may expect that
if an old SNR interacts with a GMC,  
TeV $\gamma$-ray flux becomes large 
because of the large density in the GMC.
Here we consider two cases for the emission from
old SNR-GMC interacting systems;
one comes from a shock running into the GMC,
and the other from the GMC illuminated by particles accelerated
at the SNR shock.
We consider the GMC with  the mass $M_c=10^5M_5 M_{\sun}$,
and the radius $R_c=18~{\rm pc}~M_5^{1/2}$ \citep{blitz2004}.
Assuming the spherical symmetry, the mean number density
of the GMC is $n_c=1.7\times10^2M_5^{-1/2}$~cm$^{-3}$.
We also assume that the magnetic field of the GMC is
$B_c=1~\MG~n_c^{1/2}$ \citep{crutcher1991}.
Then we find
\begin{equation}
B_c=13~\MG~M_5^{-1/4}~~,
\end{equation}
and the Alfv\'{e}n velocity in the GMC is
$v_A=2.2\times10^5$~cm~s$^{-1}$.
In this section, numerical values are for
the fiducial parameter set 
($E_{51}=v_{i,9}=n_0=h=B_{\rm ISM,-5}=M_5=1$)
unless otherwise stated.

\subsection{Emission from a shock running into a GMC}
\label{subsec:emissionGMC}

When an SNR shell collides with a GMC with the number density
$n_c$, the shock front is formed in the GMC.
In general, since the geometry of SNR shell and the GMC at the
collision is unknown, it is hard to precisely determine the dynamics
of the shell. Here we perform a very rough calculation.
Assuming the momentum conservation, 
the shell velocity, $v_{\rm s, c}$, in the GMC can be related with the
 velocity, $v_s$, of the SNR shell just before the collision as
\citep{chevalier1999}
\begin{equation}
v_{\rm s, c}\sim \f{v_s}{1+(n_c/n_{\rm sh})^{1/2}}~~. 
\end{equation}
The radiative cooling is efficient and the shock is isothermal.
We find that the velocity of the shock in the GMC is
$v_{\rm s, c}\sim1.1\times10^7$~cm~s$^{-1}$
when the SNR shell with an age of $4.6\times10^4$~yrs
($v_s=3.9\times10^7$~cm~s$^{-1}$ and 
$n_{\rm sh}=25$~cm$^{-3}$) 
collides with the GMC with the mass of $10^5M_{\sun}$.
Then, the Alfv\'{e}n Mach number of the shell running in the GMC
is $\sim50$, and  the compression ratio is estimated as
 $r\sim\sqrt{2}v_{\rm s, c}/v_A\sim70$,
so that the shock acceleration works well \citep{bykov2000}.
The downstream magnetic field is 
\begin{equation}
B_\d=(8\pi m_Hn_c)^{1/2}v_{\rm s, c}=9.1\times10^2~\MG~~,
\end{equation}
and we find $E_{\rm max,p}=79$~TeV and $E_{\rm max,e}=0.16$~TeV.
Here we adopt $p=1$ as a limiting case, because
the compression ratio is much larger than unity.
The simulated spectrum is shown in  Fig.~\ref{fig:spec_gmc}.
We assume the electron-to-proton ratio 
$K_{\rm ep}=1\times10^{-3}$ as in \S~\ref{sec:emissionSNR}, 
however, our conclusion does not
change if the ratio becomes ten times larger.
Since $E_{\rm max,e}$ is  small,
both X-ray and TeV $\gamma$-rays are hadronic origin.
Because of the large magnetic field, the secondary synchrotron radiation
is strong in the X-ray band, and
the flux ratio is not so large, $R_{\rm TeV/X}=6.9$.

We calculate $R_{\rm TeV/X}$ for several cases in which
the shocked GMC has $v_{\rm s, c}=1.1\times10^7$~cm~s$^{-1}$, 
and find that the large magnetic field at the emitting region
causes bright secondary synchrotron radiation, and
that the flux ratio for the emission from shocked GMC
does not exceed $\sim20$.
We also note that if we adopt $p=2.0$ instead of $p=1.0$,
with other parameters fiducial,
we obtain $R_{\rm TeV/X}=15$, which implies
the value of $p$ does not affect our conclusion.

If the SNR shell with $t_{\rm age}\gtrsim5\times10^4$~yrs
collides with the GMC with $M_5=1$,
$v_{\rm s, c}$ is smaller than $1.1\times10^7$~cm~s$^{-1}$,
so that the TeV emission from the shock in the GMC
is significantly weak because of
the wave damping effect and the energy loss effect
as discussed in \S~\ref{subsec:emissionSNRold}.
At this time, the density is so high that high-energy
protons lose their energy via pion production.
Hence when SNR with $t_{\rm age}\gtrsim5\times10^4$~yrs
collides with GMC, the arising shock in the GMC
does not emit TeV $\gamma$-rays.

\subsection{Emission from a GMC Illuminated by protons accelerated
at an Old SNR shock}
\label{subsec:emissionGMC2}

If an SNR  with $t_{\rm age}\gtrsim5\times10^4$~yrs
interacts with a GMC with $M_5=1$,
particle acceleration at the GMC shock is inefficient to produce
TeV $\gamma$-ray emitting particles
because $v_{\rm s, c}$ is smaller than $1.1\times10^7$~cm~s$^{-1}$.
Nevertheless, the high density of GMC  works as a target
of the high-energy protons to produce pions.
They are accelerated at the shock front of the SNR
and penetrate into the GMC, radiating photons.
We have seen in \S~\ref{subsec:emissionSNRold} that the SNR shock
itself can accelerate TeV-$\gamma$-ray-emitting particles
until $t_{\rm age}\sim3\times10^5$~yrs.
Here we consider the emission from the GMC encountered by the
SNR with $t_{\rm age}=3\times10^5$~yrs.
The simulated spectrum is shown in Fig.~\ref{fig:spec_gmc2}.
We use $E_{\rm max,p}=42$~TeV and $E_{\rm max,e}=0.58$~TeV
as seen for the fiducial case in \S~\ref{subsec:emissionSNRold}, 
while we adopt the magnetic field of the emission region
as $B_c=13$~$\MG$.
Both the X-rays and TeV $\gamma$-rays are again hadronic origin. 
Compared with Fig.~\ref{fig:spec_gmc},
the secondary synchrotron radiation
is weaker in the X-ray band because of smaller magnetic field.
We find $R_{\rm TeV/X}=5.9\times10^2$.
If the GMC is encountered by the SNR with 
$t_{\rm age}\gtrsim3\times10^5$~yrs, particles emitting TeV $\gamma$-rays
do not exist around the shock front as seen in \S~\ref{subsec:emissionSNRold}.

We calculate $R_{\rm TeV/X}$ for several cases in which
parameters are changed within a reasonable range,
and find that the value of $R_{\rm TeV/X}$ 
remains unchanged within a factor of three.
We also note that if we adopt $p=2.0$ instead of $p=1.5$,
with other parameters fiducial,
we obtain $R_{\rm TeV/X}=1.4\times10^3$, which implies
the value of $p$ does not affect our conclusion.

For $B_c=13$~$\MG$,
both TeV $\gamma$-rays and X-rays are originated in 
the accelerated protons with energies of more than 10~TeV.
Let $\delta=\delta_{\rm pc}$~pc be the separation between the GMC and
the SNR shell producing high-energy particles.
The time for a particle with an energy 
$E_{\rm cr}=10E_{\rm cr, 10TeV}$~TeV to
reach the GMC is estimated by the
diffusion time, 
$t_{\rm dif}\sim2\times10^3\eta^{-1}B_{\rm ISM,-5}
E_{\rm cr, 10TeV}^{-1}\delta_{\rm pc}{}^2$~yrs,
where $\eta$ is the gyro factor.
If $\delta_{\rm pc}\sim1$, we expect $t_{\rm dif}$ is much smaller than
the age of the SNR.
Hence all particles in the energy regime in which we are interested
reach the target GMC almost simultaneously, which implies
that the particle spectra at the GMC does not so much depend
on the character of propagations from the SNR shell to the GMC.
If $\delta_{\rm pc}\gg1$, effects of energy-dependent diffusion
should be considered. Such a detailed calculation is not
considered here.

\section{Discussions}
\label{sec:dis}

For an old SNR in the radiative phase, 
the maximum energy of primary electrons
is so small that emissions via leptonic processes are vanishingly 
small both in the X-ray and TeV $\gamma$-ray bands.
On the other hand,  
there still exists  accelerated protons with energies of
more than $\sim10$~TeV.
These particles emit the TeV $\gamma$-rays
via $\pi^0$-decay and synchrotron X-rays from secondary
electrons generated by charged pions.
We find that  the ratio
\[
 R_{\rm TeV/X}=\f{F_\gamma(1-10~{\rm TeV})}{F_X(2-10~{\rm keV})}
\]
could be more than $\sim10^2$.
Such  sources may be an origin of recently discovered
unidentified TeV sources, and give us
the evidence for hadron acceleration.
We might have to consider the interaction
between the GMC and the SNR with an age of $\sim10^5$~yrs.
For $t_{\rm age}\ll10^5$~yrs,
SNR radius is so small that there is only a few SNRs interacting
with a GMC.
On the other hand, if the SNR with  $t_{\rm age}\gg10^5$~yrs
encounters the GMC, 
there are few high-energy particles emitting TeV $\gamma$-rays
around shocks of the SNR and the GMC  because of the energy loss effect
and/or the wave damping effect occurring at low-velocity isothermal
shocks.

Actually detected TeV sources have the energy flux 
$F_\gamma(1-10~{\rm TeV})\sim10^{-12}$--$10^{-11}$erg~cm$^{-2}$s$^{-1}$
\citep{aharonian2005,aharonian2005d}.
Hence if $R_{\rm TeV/X}\la10^{2-3}$, then
$F_X(2-10~{\rm keV})\ga10^{-14}$~erg~cm$^{-2}$s$^{-1}$.
Such diffuse, extended source can be detected with current
X-ray telescopes ({\it Suzaku}, {\it Chandra}, and {\it XMM-Newton}).
Especially, X-ray Imaging Spectrometer (XIS) onboard {\it Suzaku} is
capable of observing dim,
diffuse X-ray sources with low background in the hard X-ray band.
Less energetic protons emit GeV $\gamma$-rays that might have been
detected by EGRET or may be detected by GLAST in the future.
However, for isothermal shocks, 
the spectrum of accelerated particles may be hard ($p<2$).
Indeed, some of
the newly discovered TeV sources  show 
 hard spectrum \citep{aharonian2005,aharonian2005d},
which may imply $p\lesssim2$.
In such a case, GeV emission becomes dim
as shown in Table~\ref{table2}, where
we calculate $R_{\rm TeV/GeV}$ for the case
of Figures~1--3, 5 and 6.
Radio emission is also expected.
However, as can be seen in Table~\ref{table2},
the sources with large $R_{\rm TeV/X}$ have
large $R_{\rm TeV/radio}$, so that the radio
emission may be dim.
This fact might account for a large scatter of
$R_{\rm TeV/radio}$ found by \citet{helfand2005}.
Furthermore, the source is confined in a galactic plane region
and radio emission may be obscured by  diffuse components.

\begin{table}
\centering
\begin{minipage}{140mm}
\caption{
Calculated flux ratios for various bands.}
\label{table2}
\begin{tabular}{cccc}
\hline
& $R_{\rm TeV/X}$%
\footnote{$R_{\rm TeV/X}=F_\gamma$(1--10~TeV$)/F_X$(2--10~keV)~.}
& $R_{\rm TeV/GeV}$%
\footnote{$R_{\rm TeV/GeV}=F_\gamma$(1--10~TeV$)/F_{\rm GeV}$(1--10~GeV)~.}
& $R_{\rm TeV/radio}$%
\footnote{$R_{\rm TeV/radio}=F_\gamma$(1--10~TeV$)
/F_{\rm radio}$($10^7$--$10^{11}$~Hz)~.}
\\
\hline
Fig.~1 & $7.6\times10^{-2}$ & $0.36$  & $9.8$            \\
Fig.~2 & $1.6$              & $0.25$  & $6.6$            \\
Fig.~3 & $82$               & $6.6$   & $47$             \\
Fig.~5 & $6.9$              & $74$    & $9.7\times10^3$  \\
Fig.~6 & $5.9\times10^2$    & $5.8$   & $3.5\times10^2$  \\
\hline
\end{tabular}
\end{minipage}
\end{table}

In this paper, we consider TeV $\gamma$-ray emission from
(1) a single old SNR with $t_{\rm age}\sim3\times10^5$~yrs 
(\S~\ref{subsec:emissionSNRold}),
(2) a shocked GMC collided by the old SNR with
$t_{\rm age}\sim5\times10^4$~yrs
(\S~\ref{subsec:emissionGMC}),
and
(3) an unshocked GMC illuminated by high-energy particles
arising at the shock of the old SNR with $t_{\rm age}\sim3\times10^5$~yrs
(\S~\ref{subsec:emissionGMC2}),
respectively.
As in the followings,
we can roughly estimate, for each case, 
the expected number
of observed TeV sources, $N_{\rm TeV}$,
which have observed flux larger than 3\% of the Crab flux
($6\times10^{-13}$~cm$^{-2}$s$^{-1}$ above 1~TeV) and
lie in the inner part of the galactic plane with
$-30^\circ<l<30^\circ$ \citep[cf.][]{aharonian2005,aharonian2005d}.
The number of  detected TeV sources  gives us important information  
on the energetics of old SNR emission.

Let us first consider the case~(1).
Let $E_{p}=E_{p,50}10^{50}$~ergs be the energy of accelerated
protons stored in the SNR.
Then, the observed flux of TeV $\gamma$-rays is calculated as
$F(>{\rm TeV})\sim1\times10^{-10} n_{{\rm sh,} 0.86}E_{p,50}
d_{\rm kpc}^{-2}$~cm$^{-2}$s$^{-1}$, where
$n_{\rm sh}=10^{0.86}n_{{\rm sh,} 0.86}$~cm$^{-3}$ and
$d=d_{\rm kpc}$~kpc are the density of the SNR shell
and the distance to the source, respectively.
The maximum distance to the source is  
$d_{\rm max}\sim13n_{{\rm sh,} 0.86}^{1/2}E_{p,50}^{1/2}$~kpc.
Then the volume fraction of the survey region is
\[
 \beta_1\sim\frac{(\pi/6)d_{\rm max}{}^2}{\pi(10~{\rm kpc})^2}
\sim0.30~n_{{\rm sh,} 0.86}E_{p,50}~~.
\]
Hence, for assumed SN explosion rate of 
$1\times10^{-2}r_{-2}$~yr$^{-1}$,
we obtain 
\begin{eqnarray}
N_{\rm TeV}&\sim&3\times10^3~r_{-2}\beta_1\nonumber\\
&\sim&9\times10^2~r_{-2}n_{{\rm sh,} 0.86}E_{p,50}~~.\nonumber
\end{eqnarray}

Next, we calculate $N_{\rm TeV}$ for case~(3).
In order to estimate the  number of GMCs in our Galaxy,
we adopt the mass function of GMCs, $f(M_c)\propto M_c^{-1.5}$
with the maximum and the minimum mass,
$M_{\rm max}=2\times10^7M_{\sun}$ and $M_{\rm min}=30M_{\sun}$,
respectively \citep{blitz2004}.
If the total mass of GMCs in our Galaxy is $2\times10^9M_{\sun}$,
then the number of GMCs with mass of around $M_c=10^5M_5M_{\sun}$
is about $7\times10^2M_5^{-1/2}$.
The SNR with an age of $3\times10^5$~yrs 
has a radius $R_s\sim90$~pc, so that the geometrical factor,
representing the fraction of accelerated protons
colliding the GMC and emitting $\gamma$-rays,
is estimated as $\alpha=(R_c/R_s)^2/4\sim1\times10^{-2}$.
The observed flux of TeV $\gamma$-rays is calculated as
$F(>{\rm TeV})\sim3\times10^{-11}\alpha_{-2} n_{{\rm c,}2.2}E_{p,50}
d_{\rm kpc}^{-2}$~cm$^{-2}$s$^{-1}$,
where $\alpha=\alpha_{-2}10^{-2}$, and
$n_{\rm c}=10^{2.2}n_{{\rm c,}2.2}$~cm$^{-3}$ is the density
of the GMC.
The maximum distance to the source is  
$d_{\rm max}\sim6\alpha_{-2}^{1/2}n_{{\rm c,}2.2}^{1/2}E_{p,50}^{1/2}$~kpc.
Then the volume fraction of the survey region is
\[
\beta_1\sim\frac{(\pi/6)d_{\rm max}{}^2}{\pi(10~{\rm kpc})^2}
\sim7\times10^{-2}\alpha_{-2}n_{{\rm c,}2.2}E_{p,50}~~.
\]
Assuming the volume of the galactic disk of 
$4\times10^{10}$~pc$^3$, the mean separation of GMCs is
$\ell\sim4\times10^2M_5^{1/6}$~pc.
Then, the probability that a SNR collides with a GMC is 
\[
\beta_2\sim\f{4\pi}{3}\left(\f{R_s}{\ell}\right)^3
\sim5\times10^{-2}
\ell_{400}{}^{-3}
\left(\f{R_s}{90~{\rm pc}}\right)^3~~,
\]
where $\ell=400~\ell_{400}$~pc.
Therefore, we derive
\begin{eqnarray}
N_{\rm TeV}&\sim&3\times10^3~r_{-2}\beta_1\beta_2\nonumber\\
&\sim&10~r_{-2}\alpha_{-2}n_{{\rm c,}2.2}E_{p,50}
\ell_{400}{}^{-3}
\left(\f{R_s}{90~{\rm pc}}\right)^3~~. \nonumber
\end{eqnarray}
Hence $E_{p,50}\gtrsim0.1$ is required so that $N_{\rm TeV}\gtrsim1$.

Finally, we consider the case~(2).
We assume that  the energy of accelerated 
protons is $1\times10^{48}E'_{p,48}$~ergs, which
is much smaller than the thermal energy stored
in the shocked GMC.
Then, the observed  TeV $\gamma$-ray flux is calculated as
$F(>{\rm TeV})\sim3\times10^{-9}n_{{\rm c,d,}4.1}E'_{p,48}
d_{\rm kpc}^{-2}$~cm$^{-2}$s$^{-1}$, where
$n_{\rm c,d,}=10^{4.1}n_{{\rm c,d,}4.1}$~cm$^{-3}$ is the density
of shocked GMC.
The maximum distance to the source is calculated as
$d_{\rm max}\sim66n_{{\rm c,d,}4.1}^{1/2}E'_{p,48}{}^{1/2}$~kpc, 
which is larger than the size of the Galactic disk.
Therefore, the volume fraction is  about $\beta_1\sim0.3$.
Since the SNR with an age of $5\times10^4$~yrs 
has a radius $R_s\sim40$~pc,
the collision probability is
\[
\beta_2\sim\f{4\pi}{3}\left(\f{R_s}{\ell}\right)^3
\sim4\times10^{-3}
\ell_{400}{}^{-3}
\left(\f{R_s}{40~{\rm pc}}\right)^3~~.
\]
Hence we obtain
\begin{eqnarray}
N_{\rm TeV}&\sim&5\times10^2~r_{-2}\beta_1\beta_2\nonumber\\
&\sim&0.6~r_{-2}(\beta_1/0.3)\ell_{400}{}^{-3}
\left(\f{R_s}{40~{\rm pc}}\right)^3~~.\nonumber
\end{eqnarray}
Taking into account for the uncertainties for $R_s$ and/or $\ell_{400}$,
we can expect $N_{\rm TeV}\gtrsim1$ for reasonable values
of parameters.

At present, the number of unidentified TeV sources showing
large $R_{\rm TeV/X}$ is only a few.
If the number of actually detected TeV sources does not increase
(i.e., the H.E.S.S. galactic plane survey with the flux
larger than 0.03~Crab above 1~TeV is nearly complete at present),
then  $E_{p,50}\lesssim0.01$ is implied, 
otherwise $N_{\rm TeV}$ for the case (1) is much larger than a few. 
This might suggest the escape of high-energy particles from the SNR
starts by  $t_{\rm age}\sim10^5$~yrs, or
injection efficiency is low for old SNRs.
Then, the number of TeV sources, $N_{\rm TeV}$, for the case (3)
should be small.
On the other hand, 
the case (2) may be still likely because the number of sources
for this case can be comparable to that for the case (1).
Hence, some TeV sources are associated with the GMC, but others
not.
In order to clarify whether the TeV source is associated with
a GMC, the CO observation may be important.

\section*{Acknowledgments}

We wish to thank the anonymous referee, J.~S.~Hiraga,
K.~Ioka, A.~Loeb, R.~Narayan, P.~Slane, 
T.~Takahashi, T.~Tanimori, and Y.~Uchiyama  for their useful comments.
K.K. was supported by NSF grant AST 0307433.
R.Y. and A.B. are
 supported in part by the Grant-in-Aid for young Scientists (B)
of the Ministry of Education, Culture, Sports, Science and
Technology (No.~18740153 and 17740183).

%



\label{lastpage}
\end{document}